\documentclass{article}
\usepackage[utf8]{inputenc}

\usepackage[a4paper, margin=0.8in]{geometry}
\usepackage[style=nature,doi=false,isbn=false,eprint=false,url=false,maxbibnames=99, ]{biblatex}
\DeclareFieldFormat{sentencecase}{\MakeSentenceCase{#1}}

\renewbibmacro*{title}{%
  \ifthenelse{\iffieldundef{title}\AND\iffieldundef{subtitle}}
    {}
    {\ifthenelse{\ifentrytype{article}\OR\ifentrytype{inbook}%
      \OR\ifentrytype{incollection}\OR\ifentrytype{inproceedings}%
      \OR\ifentrytype{inreference}}
      {\printtext[title]{%
        \printfield[sentencecase]{title}%
        \setunit{\subtitlepunct}%
        \printfield[sentencecase]{subtitle}}}%
      {\printtext[title]{%
        \printfield[titlecase]{title}%
        \setunit{\subtitlepunct}%
        \printfield[titlecase]{subtitle}}}%
     \newunit}%
  \printfield{titleaddon}}
\AtEveryBibitem{\clearfield{month}}
\AtEveryCitekey{\clearfield{month}}
\AtEveryBibitem{\clearlist{language}}
\AtEveryBibitem{\clearlist{urldate}}

\DeclareUnicodeCharacter{2212}{-}

\usepackage{siunitx}
\usepackage{tabularx}[]
\usepackage{array}
\usepackage{amsmath}
\usepackage{amssymb}
\usepackage{booktabs}
\usepackage{makecell}
\usepackage{enumitem}
\usepackage{graphicx}
\usepackage{caption}
\usepackage{subcaption}
\usepackage{float}
\usepackage{hyperref}
\usepackage{url}
\usepackage[protrusion=true]{microtype}
\usepackage[capitalise,noabbrev]{cleveref}
\usepackage{fancyhdr}
\usepackage{authblk}
\usepackage[symbol]{footmisc}

\title{Confronting the thermodynamics knowledge gap: A short course on computational thermodynamics in Julia}
\author[1,2]{Luc T. Paoli}
\author[1,2]{Pavan K. Inguva}
\author[1]{Andrew J. Haslam*$^{,}$}
\author[1,3]{Pierre J. Walker*$^{,}$}
\affil[1]{Department of Chemical Engineering, Imperial College London, London SW7 2AZ, United Kingdom}
\affil[2]{Massachusetts Institute of Technology, 77 Massachusetts Avenue, Cambridge, MA 02139, United States}
\affil[3]{Division of Chemistry and Chemical Engineering, California Institute of Technology, Pasedena, CA 91125, United States}

\date{}

\begin{document}

\maketitle
\footnotetext[1]{Corresponding authors. Email: a.haslam@imperial.ac.uk, pjwalker@caltech.edu}
\renewcommand{\thefootnote}{\arabic{footnote}}

\begin{abstract}

Computational elements in thermodynamics have become increasingly important in contemporary chemical-engineering research and practice. However, traditional thermodynamics instruction provides little exposure to computational thermodynamics, leaving students ill-equipped to engage with the state-of-the-art deployed in industry and academia. The recent rise of easy-to-use open-source thermodynamic codes presents an opportunity for educators to help bridge this gap. In this work, we present a short course that was developed and rolled-out using the \texttt{Clapeyron.jl} package, the material of which is all openly available on GitHub. The course can serve as a foundation for others to similarly integrate computational material in thermodynamics education. The course is structured into three sections. Section one serves as a refresher and covers core material in numerical methods and thermodynamics. Section two introduces a range of thermodynamic models such as activity-coefficient models and cubic equations of state, outlining their implementation. In section three the focus is moved to deployment, guiding students on how to implement computational-thermodynamics methods covering volume solvers, saturation solvers, chemical-stability analysis and flash problems. In a pilot study conducted with both undergraduate and graduate students, participants found the material engaging and highly relevant to their chemical-engineering education. 

\end{abstract}

\section{Introduction}

The application of thermodynamics in industry and research is becoming increasingly diverse and sophisticated with advances in computational capabilities, available resources, theoretical understanding, and use-cases \cite{hendriks_industrial_2010,kontogeorgis_industrial_2021,kontogeorgis_conclusions_2022,de_hemptinne_view_2022}. The expanding role that thermodynamics plays in the progress of science and engineering is not surprising considering its centrality in describing physical systems. Nevertheless, it places a challenge on educators to provide suitably advanced thermodynamics instruction to ensure students are well-equipped to meet their professional needs after graduation. Accordingly, the  continued development of education opportunities and resources for thermodynamics remains as important as ever. 

One of the most common tasks encountered is the use of various thermodynamic models and methods to analyze systems of interest, to generate estimates for various properties (e.g., phase diagrams or activity coefficients), or as part of a process / physics-based simulation at a larger scale. In that regard, there are several currently available software packages of varying degrees of functionality that have been used in professional and/or teaching settings e.g., ASPEN \cite{sandler_using_2015}, PYroMAT \cite{martin_pyromat_2022,martin_problem-based_2017}, CoolProp \cite{bell_pure_2014}, XSEOS \cite{castier_xseos_2011}, teqp \cite{bell_implementing_2022} , and custom code \cite{martin_teaching_2011}. 
A key component that is implicit in the use of these models and software packages is the computational aspect of thermodynamics, yet this is often neglected in thermodynamics courses, even at a graduate level. 
Often, one of the major selling points of many software packages is the abstraction of the numerical implementation of the model and algorithms, which can be complex, thus enabling users to rapidly employ advanced thermodynamic models in their workflow (e.g., see \cite{walker_clapeyronjl_2022,bell_implementing_2022}).
 
In conceptualizing an advanced teaching course, it is imperative to not treat the various models and packages as ``black boxes" which ``magically" generate the desired answer \cite{karimi_use_2005,martin_teaching_2016}, as the objective of such advanced courses is to help students develop deeper insights and become informed practitioners. Therefore, course content and packages employed should not only allow students to engage meaningfully with the theory and use of various thermodynamic models (including state-of-the-art models where practicable), they should also provide the opportunity to explore various back-end details such as the numerics and algorithms needed to compute the desired result. Other considerations that should be taken into account in relation to software for use in educational settings include availability and ease-of-use \cite{martin_teaching_2016,inguva_introducing_2021,inguva_how_2021}.

In this work, we outline the development and roll-out of a short course for introducing computational thermodynamics to advanced undergraduate / graduate students. The course is intended to be a stand-alone course that can be delivered on an ad-hoc basis, though we intend to integrate components of the course into a graduate thermodynamics module. The course was designed to be modular so that self-directed learners and/or educators can adapt specific components for their own needs or develop an expanded course using the material presented as a foundation. All course material and code are currently publicly available in a GitHub repository and additional contributions are planned to expand the material offered.  

The remainder of this article is laid out as follows: the context in which the course is / can be taught is provided in section \ref{sect:course_context}. After a short discussion on the computational environment (section \ref{sect:coding}), the structure of the course is provided in section \ref{sect:course_struct}. The results and feedback obtained from a pilot run of the course is presented and discussed in section \ref{sect:results}. Finally, the overall findings of this article are summarised in section \ref{sect:conclusion}. 

\section{Course Context}
\label{sect:course_context}

The short course was developed to serve a dual role of introducing advanced undergraduate students and graduate students to advanced thermodynamic modelling tools and real-world applications of the statistical thermodynamic theories. Students taking this course should already be familiar with the fundamentals of thermodynamics (for example, the laws of thermodynamics, Maxwell relations, and phase equilibria) and, perhaps, some thermodynamic models (such as the ideal gas, van der Waals equation of state, and ideal-solution models), as well as having taken some form of introductory coding class. The latter is not necessarily a strong requirement as the course material is delivered in such a way that the students only do minimal coding. Within the context of the Imperial College undergraduate degree program, by the end of the second year, students will have taken two courses on thermodynamics, covering all of the topics mentioned previously, and one introductory course on MATLAB coding, which we deem more than sufficient for a student intending to take part in this course.

For the benefit of those students who are not familiar with all of the above pre-requisites or
% , in the case of graduate students, need to be re-familiarised with some material, 
require a refresher for any of the material, the first section of the course provides sufficient introductory content. Conversely, to suit participants who are already familiar with large sections of this course, or instructors who wish to use only specific sections in their own class, the course is designed to be modular, allowing the user to engage with the material flexibly.

\section{Computational Environment}
\label{sect:coding}
The main software package used in this course is the recently developed open-source fluid thermodynamics toolkit Claperyon.jl \cite{walker_clapeyronjl_2022} which is implemented in the Julia language. The release of Claperyon.jl has enabled users to employ a large variety of thermodynamic models (ranging from activity-coefficient models and standard cubics to the state-of-the-art SAFT equations) in an easy-to-use and extensible manner. The package has several built-in capabilities for thermophysical-property estimation and is currently under active further development to incorporate additional capabilities to make it the go-to thermodynamics toolkit. Claperyon.jl has garnered significant interest from a diverse group of users with over 135 stars on GitHub. 

Although the Julia language is less mature than alternatives such as Python and MATLAB, it has several features and packages that make it appealing for scientific computing while also being well-suited in an educational setting \cite{pawar_cfd_2019,boukouvala_computational_2020}. In terms of ease-of-use, Julia has a similar syntax to MATLAB and Python which helps reduce challenges students may have with reading and writing code \cite{perkel_julia_2019}. Julia also has native package-management tools that can negate the need for setting up environments, reducing time spent introducing students to the coding environment. 

\begin{figure}
    \centering
    \includegraphics[width=0.6\textwidth]{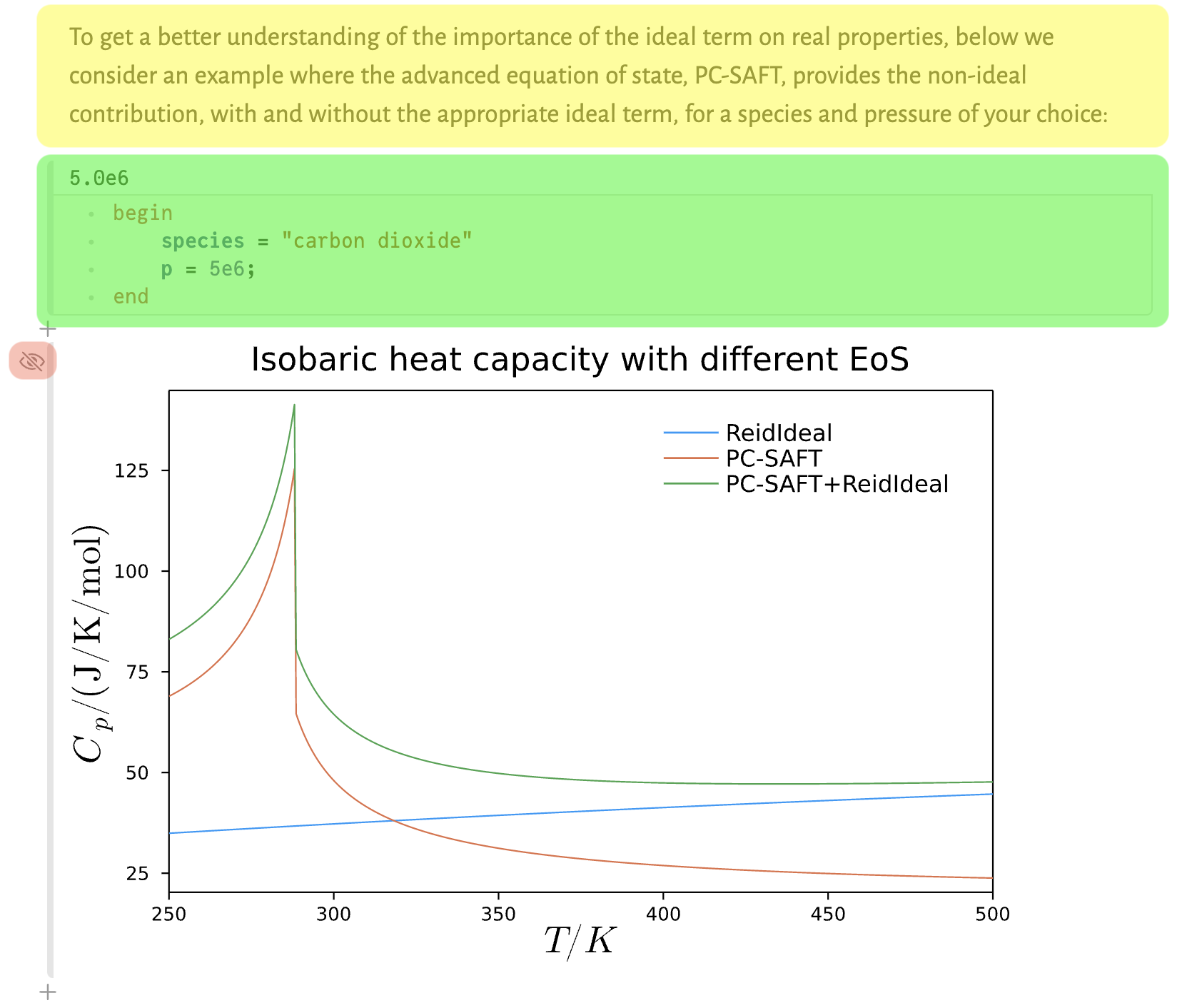}
    \caption{Example taken from the Pluto notebook used in the course. Explanatory text is here highlighted in yellow; example code is highlighted in green. To reveal the hidden code used to generate the embedded chart, students would simply select the symbol highlighted in red. The  chart represents the isobaric heat capacity of carbon dioxide obtained using different equations of state.\cite{reid_properties_1987,gross_perturbed-chain_2001}. Students are invited to vary which species and equation of state is used to examine how the ideal contribution plays a role in determining real properties.}
    \label{fig:pluto}
\end{figure}

One particular package we would like to highlight is Pluto.jl \cite{van_der_plas_plutojl_2023}, which provides \textit{reactive} notebooks. For the presentation of a course developed previously on introducing partial-differential-equation solvers\cite{inguva_introducing_2021}, a combination of ready-made Python scripts for code and accompanying pdf files was used for the course material. From student feedback, two main challenges were noted: some students experienced difficulties in setting up the Python environment needed to run the scripts; and the lack of active and structured learning activities hampered student engagement with the material. Pluto.jl can elegantly resolve both these issues. For this course, we decided to integrate both the lecture notes and codes for a given section within a single file. Illustrated visually in figure \ref{fig:pluto}, within Pluto notebooks, we are able to provide both the code needed to generate the results as well as the text explaining these results. As seen in figure \ref{fig:pluto}, the explanatory text along with the main function call and plotting code are embedded seamlessly. Pluto has the advantage of avoiding clutter in the notebook by allowing some sections of the code (e.g., plotting) to be hidden; hidden text can be revealed if desired by selecting the designated symbol (highlighted in red in figure \ref{fig:pluto}).

Although similar functionalities are available in more commonly used \textit{interactive} environments such as Jupyter notebooks, which have already found widespread use in education\cite{johnson_benefits_2020,dominguez_teaching_2021,bascunana_impact_2023}, the reactive nature of Pluto has one major additional benefit to users and educators. In the example shown in figure \ref{fig:pluto}, students are invited to modify the provided code and re-run the cell. In doing so, the entire notebook `reacts' to update every other cell block, including the hidden plotting code. As a result, the figure will automatically update based on the changes they have made. Interactive notebooks by virtue of their sequential nature do not allow for such dynamic behaviour \cite{perkel_reactive_2021}. We believe reactive notebooks allow students to engage more intently with the lecture material and, thereby, foster independent learning\cite{kalogeropoulos_facilitating_2020}. Further details on setting up the computational environment for this course are provided in the online repository.

\section{Course Structure}
\label{sect:course_struct}

\begin{figure}[!ht]
    \centering
    \includegraphics[width=0.95\textwidth]{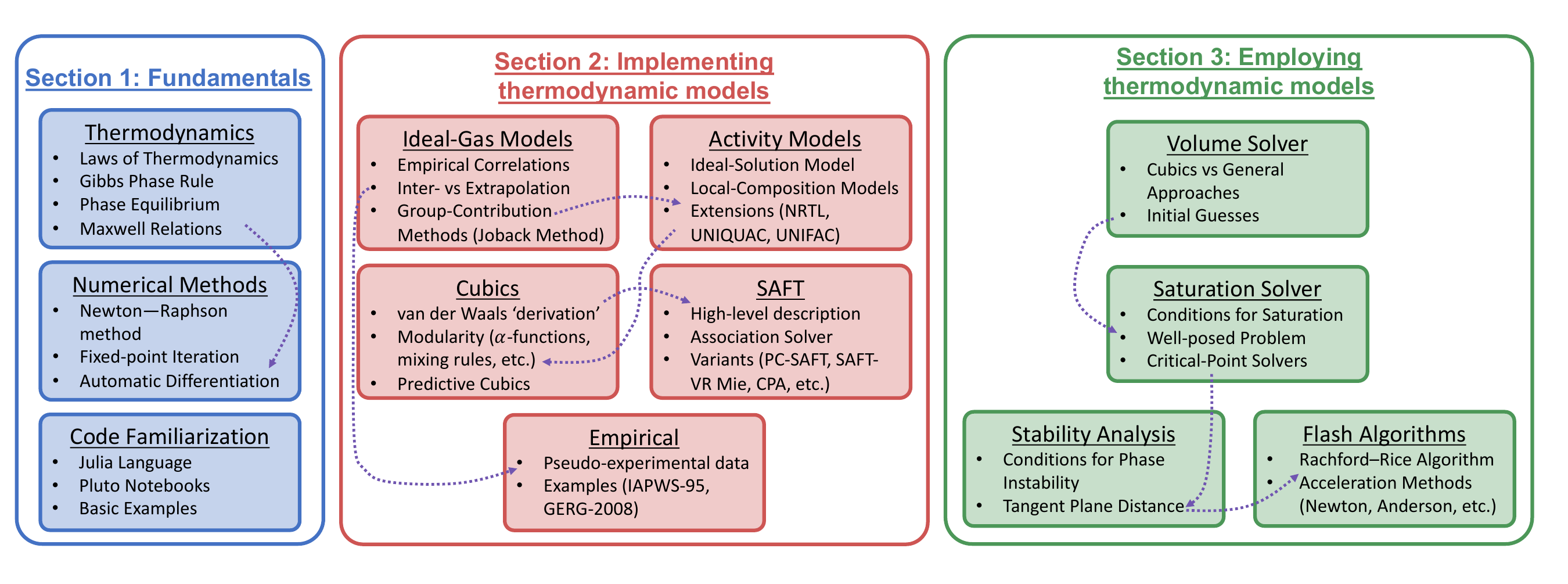}
    \caption{Course structure summary with bullets providing a brief description of each subsection. Dotted purple arrows denote concepts carried over between subsections of the course.}
    \label{fig:course_struct}
\end{figure}

Progression was used as the guiding principle when designing the course, to help accelerate students' learning in a short period of time \cite{xie_accelerating_2020}. The course is divided into three sections, each of which is delivered in a three-hour lecture (including two ten-minute breaks). An outline of the course structure is presented in figure~\ref{fig:course_struct}. The first section serves as an introduction and provides a refresher on the fundamentals of thermodynamics; important numerical methods (such as Newton's algorithm and automatic differentiation) are outlined. In the second section  five classes of equations of state are examined, providing students with a high-level overview of each class, as well as their respective advantages and disadvantages. The final section builds on material from the previous two sections and examines various computational thermodynamic problems, such as rooting-finding algorithms, stability analysis and flash algorithms. Details on theory and numerical methods for each problem are also covered.

\subsection{Section 1: Fundamentals}
The first section of the course is intended to serve as a review to ensure participants are familiar with the pre-requisites of the course. Two main topics are covered: thermodynamics; and numerical methods.  In the thermodynamics review, we revisit fundamental concepts such as thermodynamic state variables (temperature, pressure, volume, number of moles etc.) and state functions. We emphasise the point that from a state function expressed in terms of three state variables, one can obtain a complete thermodynamic description of the system using derivatives (e.g., Maxwell relations).

Having set the context for how thermodynamic properties can be generated, the core numerical methods that will be utilised are then briefly covered. We first introduce automatic differentiation, which can be used to obtain numerically exact derivatives of arbitrary functions. Its use in many computational applications, including machine learning, is highlighted to students. 
% as we believe having an appreciation of automatic differentiation is of value. 
Subsequently, root-finding algorithms (mainly Newton--Raphson and fixed-point iterations) and optimisation are discussed. At this stage, students are expected only to be aware of the existence of these methods, how they are implemented (conceptually), and their advantages and limitations. Students should recognise that errors arising during the implementation of thermodynamic models and methods could originate from the numerical methods used and not just from a limitation in their thermodynamics knowledge.

With the exception of a short exercise on how to use automatic differentiation in Julia, this section of the course was delivered through PowerPoint slides only. The remainder of the allocated lecture time was spent ensuring all students had working notebooks, to allow them to actively take part in the remaining two sections. 

% \begin{figure}[!ht]
%   \centering
%     \includegraphics[width=0.5\textwidth]{Figures/exercises/ideal_real.pdf}
%     \caption{Isobaric heat capacity of carbon dioxide obtained using different equations of state.\cite{reid_properties_1987,gross_perturbed-chain_2001}. Students are invited to vary which species and equation of state is used to examine how the ideal contribution plays a role in determining real properties.}
%     \label{fig:ideal_real}
% \end{figure}

\subsection{Section 2: Implementing thermodynamic models}

In this section fives classes of thermodynamic models are discussed; these are intended to cover the range of models that are typically encountered in academia or industry. The order in which they are taught was chosen so as to start with more-familiar approaches (ideal gas and ideal solution) and gradually approach more-advanced models such as SAFT and empirical equations of state.

\subsubsection{Ideal-gas equation}
\label{sect:ideal}

We felt that students would be most comfortable starting with something with which they should all be familiar; the obvious choice is therefore the ideal-gas equation,
% something with which all students should be familiar, we provide a physical picture an ideal gas.
% : infinitesimally small particles experiencing perfectly elastic collisions. 
% However, we quickly highlight that 
% the familiar ideal-gas equation,
%
\begin{equation}
    pV=nRT,
    \label{eq:ideal}
\end{equation}
where $p$ is pressure, $V$ is volume, $n$ is the number of moles, $R$ is the universal gas constant and $T$ is temperature, we quickly highlight that the equation alone does not constitute the full picture of an ideal gas. By integrating \eqref{eq:ideal} with respect to volume in order to obtain the Helmholtz free energy, $A$,
\begin{equation}
    A=-\int pdV = -nRT\ln{V}+c(n,T),
\end{equation}
we notice that there are missing contributions corresponding to the translational, rotational and vibrational modes of motion. These contributions are vital when obtaining real properties such as heat capacities and speeds of sound, particularly in the gas phase \cite{walker_new_2020}. We illustrate that these missing contributions can be obtained using ideal-heat-capacity correlations, with which students may already be familiar from mass- and energy-balances. The importance of these contributions is highlighted in the case of real gases (shown in the embedded chart in figure \ref{fig:pluto}). Although there are no exercises in this section, students are invited to vary the species and conditions for which this chart is generated in order to appreciate the importance of these contributions.

Enabled by the simplicity of the ideal-gas model, we take the opportunity to introduce students to other important concepts in computational thermodynamics. For instance, while correlations provide simple models to address typically complex properties, one does need to be careful to only use these for \textit{interpolation} as the accuracy of these models can degrade quickly when \textit{extrapolating}. We also introduce students to the concept of group-contribution approaches, which can be very useful when trying to model species for which no (molecular) parameters exist. While such approaches are more common in relation to other models discussed later in the course, it is helpful to first introduce these approaches in the context of a more-familiar model. 

\subsubsection{Activity coefficient models}

Although some students may already be familiar with the ideal-solution model, we begin this section by describing the assumptions implicit in this approach, and how they lead to the more-familiar Raoult's law,
\begin{equation}
    p_i=x_ip_{\mathrm{sat},i}\,
\end{equation}
where $p_i$, $p_{\mathrm{sat},i}$ and $x_i$ are the partial pressure, saturation pressure and molar composition of species $i$, respectively. It is then easy to introduce the concept of activity coefficients ($\gamma_i$) as deviations from the ideal-solution model, motivating the interest in activity-coefficient models. We also highlight some of the limitations of activity-coefficient models, primarily being that they can only be used to obtain mixture properties. 

Having now established the ideal-solution model, it is then simple to explain the local-composition model, from which all the activity models we present in this section are derived. We use the Wilson model\cite{wilson_vapor-liquid_1964} where the key interaction parameter, $\Lambda_{ij}$, between two species $i$ and $j$ is given by:
\begin{equation}
    \Lambda_{ij} = \frac{v_j}{v_i}\exp{-\Delta \lambda_{ij}/(RT)}\,
\end{equation}
where $v_i$ is the molar volume of species $i$ and $\Delta \lambda_{ij}$ is the interaction energy between $i$ and $j$. We then highlight the two primary effects captured by local-composition models: size effects ($v_j/v_i$) and enthalpic effects ($\Delta \lambda_{ij}$). As a result, it is straightforward to explain the assumptions that underly the other models (NRTL\cite{renon_local_1968} and UNIQUAC\cite{abrams_statistical_1975}) discussed in this section, allowing us to focus on the advantages and limitations of each. One of the key messages we highlight is that, to decide which of these approaches is `best' for a given application, one needs to compare the predictions to experimental data. 

Finally, as we have already introduced the concept of group-contribution approaches in relation to ideal-gas models, it is straightforward to now introduce the student to perhaps the most-commonly used activity coefficient model: UNIFAC\cite{fredenslund_computerized_1977}. We also highlight another issue students may encounter when using thermodynamic models, particularly UNIFAC: differences in models used by specific individuals but often referred to by the same name, 
% The example we give is how there are
exemplified here by the two versions of UNIFAC (the `original' and Dortmund versions), which are both referred to by the same name in literature. We emphasise the importance of verifying the source of parameters, as well as their predictive accuracy.

\subsubsection{Cubic equations of state}
\begin{figure}
    \centering
    \includegraphics[width=0.5\textwidth]{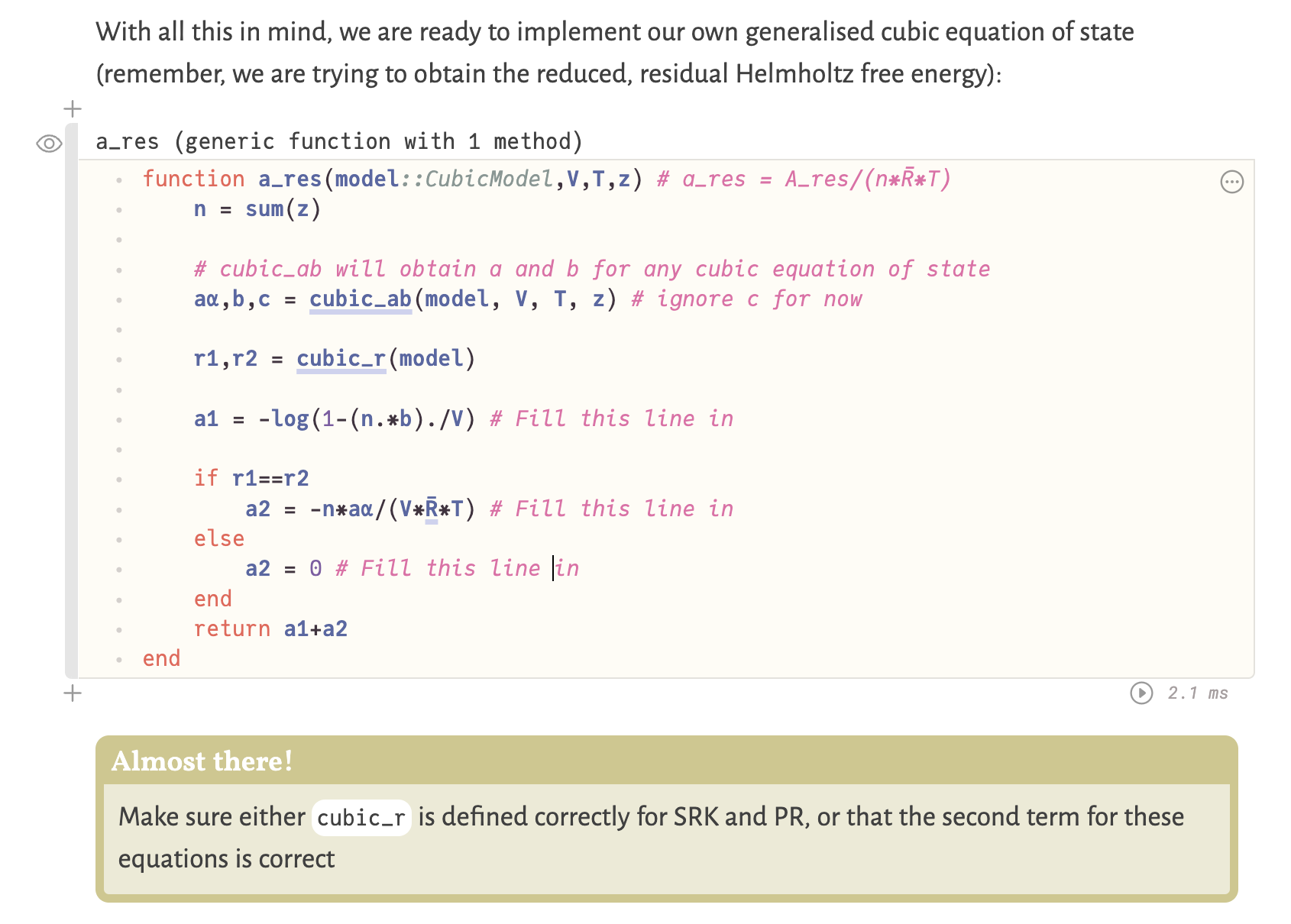}
    \caption{Example of exercise given in the cubic equation of state section. Each student is tasked with completing the code at the commented sections. A textbox below the code block will dynamically update based on the student's progress.}
    \label{fig:cubic_example}
\end{figure}

As the most-common class of thermodynamic models used in industry, we dedicate a significant portion of section two to cubic equations of state. To give some physical insight behind cubic equations of state, we
% `derive' the van der Waals equation\cite{van_der_waals_molekulartheorie_1890} from a phenomenological perspective. We also highlight how the
begin with a phenomenological discussion of the van der Waals equation\cite{van_der_waals_molekulartheorie_1890}, and remind students how the parameters can be obtained from the critical pressure and temperature of any species, allowing for a wide range of applicability. 
From here, we highlight the limitations of the van der Waals equation and how the subsequent cubic equations of state represent improvements.
% upon the van der Waals equation. 
For conciseness, we introduce only the Redlich--Kwong (RK)\cite{redlich_thermodynamics_1949}, Soave-Redlich--Kwong (SRK)\cite{soave_equilibrium_1972}, and Peng--Robinson (PR)\cite{peng_new_1976} equations but do point out that many others have been developed. In the case of the latter two equations of state, we also introduce the concept of an $\alpha$-function and its relationship to the acentric factor ($\omega$). We then emphasise one of the key advantages of cubic equations of state: they have a simple and universal form,
\begin{equation}
    p = \frac{nRT}{V-nb} - \frac{n^2a\alpha(T)}{(V+r_1nb)(V+r_2nb)}\,,
    \label{eq:cubic}
\end{equation}
where $\alpha(T)$ is the $\alpha$-function and, $a$ and $b$ are substance-specific parameters. The coefficients $r_1$ and $r_2$ are equation of state-specific. We then give students the opportunity to implement their own generalised cubic equation of state. We provide them with a code template with parts left incomplete (shown in figure \ref{fig:cubic_example}). They are then asked to define the coefficients $r_1$ and $r_2$ for the van der Waals, SRK and PR equations of state and, write out the residual Helmholtz free energy corresponding to equation \ref{eq:cubic}, accounting for each equation of state. Aside from just gaining experience in coding, there are implicit learning objectives to this exercise. In implementing a generalised cubic equation of state, the students make use of the multiple-dispatch feature in Julia and, in order to ensure that the automatic derivatives of the Helmholtz free energy are correct, they will have to explicitly define all the state variables (an easy mistake to make whenever implementing any equation of state). A text box is added below the code block which provides real-time feedback based on their current implementation and provides hints based on what part of their code is incorrect. This feature arises from the reactive nature of Pluto notebooks.

Subsequently, we illustrate one of the key benefits of cubic equations of state: modularity. We highlight how one can improve the modelling capabilities of cubic equations of state by replacing the $\alpha$-functions and adding a volume-translation correction. The impact on real properties is also illustrated with multiple examples. 

\begin{figure}
    \centering
    \includegraphics[width=0.5\textwidth]{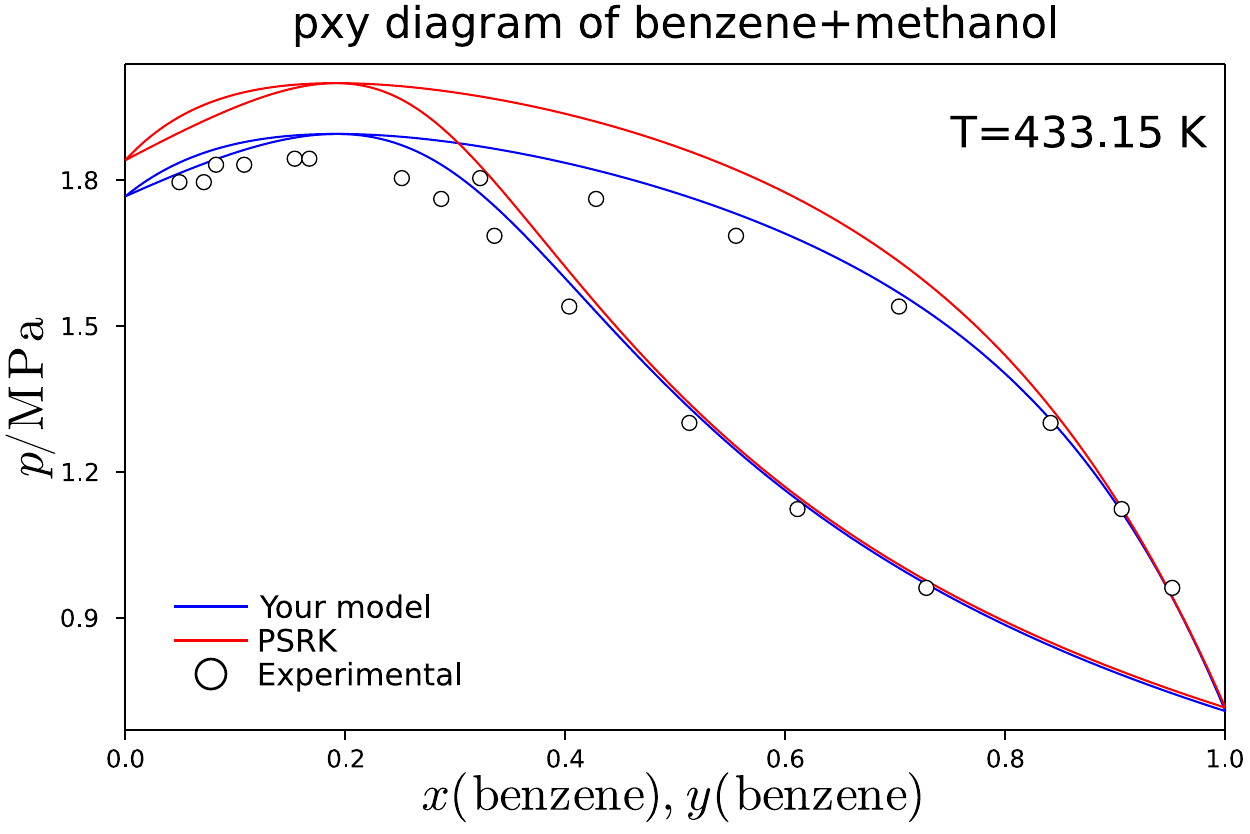}
    \caption{Pressure--composition ($p,xy$) diagram of benzene+methanol at 433.15\,K obtained using PSRK and a custom cubic equation of state.\cite{holderbaum_psrk_1991,butcher_thermodynamic_2007} Each student is invited to attempt to surpass the accuracy of PSRK by developing their own cubic equation of state, varying the $\alpha$-function, mixing rule and activity coefficient model.}
    \label{fig:your_cubic}
\end{figure}

Having covered how one can use cubic equations of state to model pure species, we now introduce students to the concept of mixing rules to model mixtures of species. We illustrate how these mixing rules map the pure-component parameters to mixture parameters in order to represent the mixture. We cover two classes of mixing rules. The first is the standard van der Waals one-fluid mixing rule, which is perhaps the most common one used in industry. It is emphasised that this mixing rule is generally inaccurate without introducing a binary interaction parameter and that these typically need to be fitted by adjustment using experimental data. The second class of mixing rule we introduce is that of the EoS/$G^E$ mixing rules. The previously-established effectiveness of activity models in modelling mixture phase equilibrium is used to motivate the concept of mixing rules which `pair' an activity-coefficient model with a cubic equation of state (such as the Huron--Vidal\cite{huron_new_1979} and Wong--Sandler\cite{wong_theoretically_1992} mixing rules).

Finally, uniting all of the previous concepts, we introduce the final and most-modern class of cubic equations of state, the predictive cubics\cite{holderbaum_psrk_1991,ahlers_development_2001}. These equations of state combine everything the students will have encountered up until now: $\alpha$-functions; volume-translation corrections; and mixing rules. We highlight that these represent the gold-standard for cubic equation of state modelling in industry. However, to emphasise the importance of validating these models against experimental data, we challenge students to try to develop their own cubic equation of state, using all the tools they have learnt up until now, to improve upon an existing predictive cubic equation of state. As shown in figure \ref{fig:your_cubic}, obtaining a better model is not necessarily challenging with the tools students have at their disposal. Note that this figure and, indeed, all the subsequent charts presented in this and the following section, is shown here exactly as seen by the students in the Pluto notebook.

\begin{figure*}[!ht]
\centering
    \includegraphics[width=0.7\textwidth]{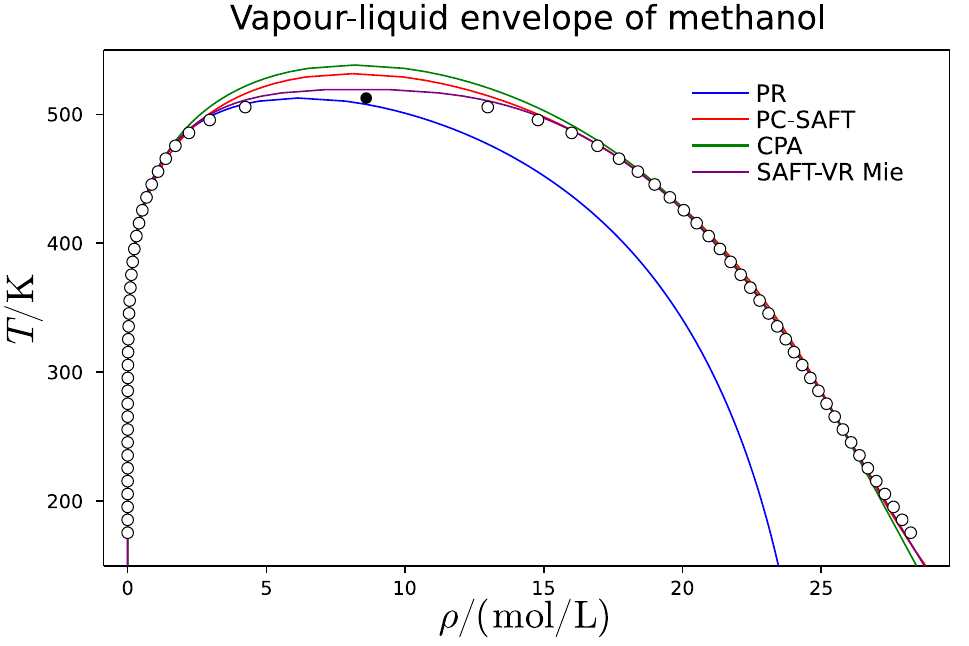}
    \caption{Vapour--liquid envelope of methanol obtained using different equations of state\cite{peng_new_1976,kontogeorgis_equation_1996,gross_perturbed-chain_2001,lafitte_accurate_2013,kunz_gerg-2008_2012}, demonstrating the significant improve of SAFT equations over cubics, but the minimal difference between individual SAFT-type equations.}
      \label{fig:saft_example}
\end{figure*}

\subsubsection{SAFT equations of state}

The Statistical Associating Fluid Theory (SAFT)\cite{chapman_saft_1989,chapman_new_1990} is not typically covered at any level within chemical-engineering curricula, primarily due to do the complexity of its derivation. However, in the past few years, SAFT equations of state have been shown to describe experimental data more accurately for a larger range of properties and species than cubic equations of state. As part of this course, we introduce students to SAFT equations of state by providing a very high-level description of how these equations represent species. Starting from the van der Waals equation, we highlight the improvements made with each term added within the SAFT equation. Before going into the individual SAFT variants, we summarise the overall SAFT approach. Some details are provided on the implementation of SAFT equations, primarily the association term which requires special treatment in order to be used. We believe that a simple, high-level understanding is sufficient for students to confidently use SAFT equations in their own work; should they want to find out more, we provide references and invite them to take an in-depth statistical-thermodynamics course provided within the college.

The remainder of this section is focussed on introducing the more-common SAFT equations of state, explaining their differences at a high level and giving their respective advantages and disadvantages. We emphasise that, as with the activity-coefficient models, to decide which approach is best, one needs to compare predictions to experimental data. However, in the case of SAFT equations, as shown in figure \ref{fig:saft_example}, a lot of the model predictions are of comparable accuracy. As a final point, we add that there is another factor to take into account when considering whether to use a SAFT equation: with their increased complexity, the computational cost of these approaches becomes significant, meaning that the students would need to weigh the added accuracy of these models with this additional cost. 

\subsubsection{Empirical equations of state}

The last class of equations of state we cover in this course is empirical equations of state. Like the ideal-heat-capacity correlations introduced in section \ref{sect:ideal}, these approaches are purely intended for interpolations. However, unlike the approaches discussed previously, while limited to relatively few systems, these equations are extremely accurate for the range of conditions in which they were regressed. As such, if students ever need to model a substance for which there already is an empirical equation of state, they might be better off using this, instead of one of the approaches discussed hitherto. As there is a large range of these models, we only provide two examples: IAPWS-95\cite{wagner_international_2008}, for water, and GERG-2008\cite{kunz_gerg-2008_2012}, for liquified natural-gas systems. We use these to illustrate the only real limitation of these models: extrapolation beyond the conditions over which they were parameterised. In short, this subsection is intended to make students aware that these empirical equations of state exist and in what cases they might wish to use them.

\subsection{Section 3: Employing thermodynamic models}

Having now developed a good understanding of the equations of state and thermodynamic models, in this section students are introduced to the computational methods used to obtain more-familiar and useful properties. One of the changes made in content presentation and delivery in this section, relative to section two, is that larger code blocks are left visible for the students to examine, given that the focus is now on the numerical implementation rather than just a high-level understanding.

\subsubsection{Volume Solvers}

\begin{figure*}[!ht]
\centering
  \begin{subfigure}[b]{0.46\textwidth}
    \includegraphics[width=1.\textwidth]{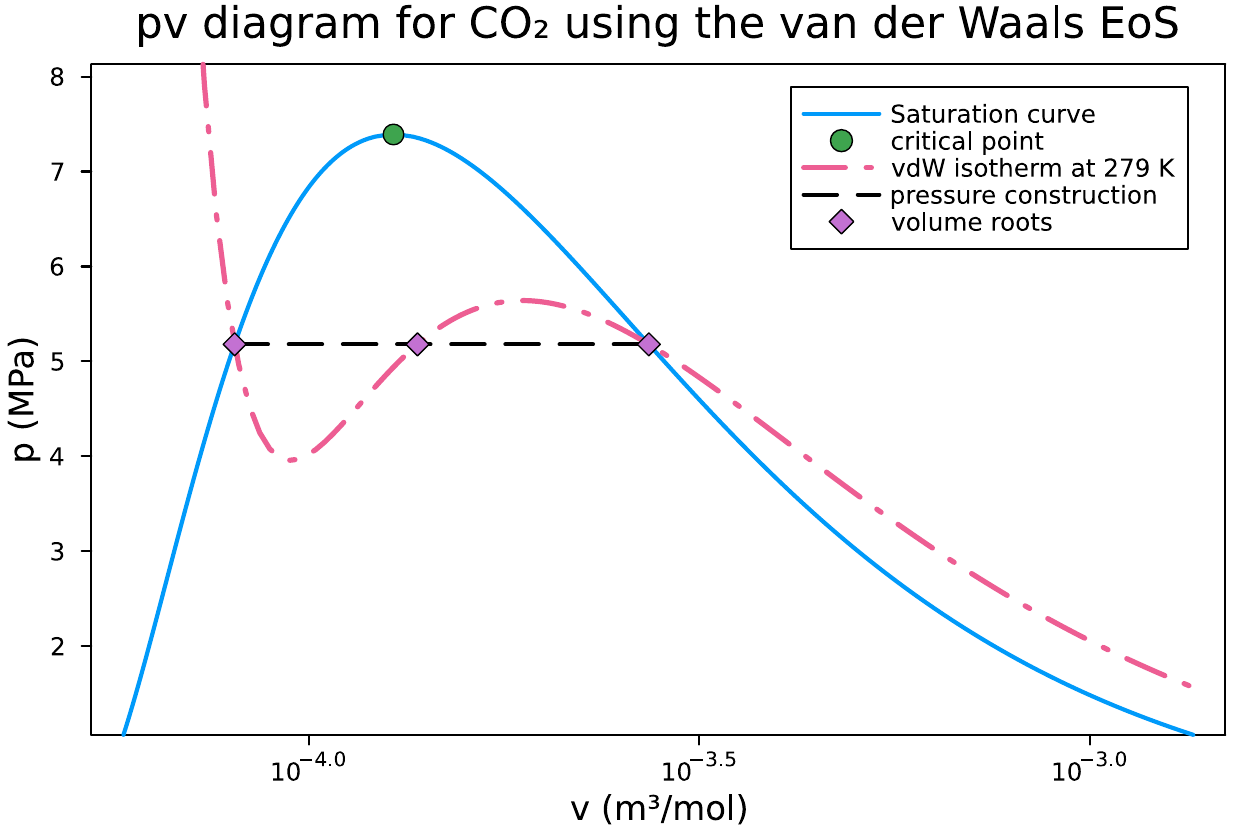}
    \caption{Example of a $(p,V)$ isotherm (pink dot-dashed line) with three possible roots (purple markers). The pressure construction is denoted by the black dashed line. The vapour--liquid envelope (blue continuous curve) and critical point (green marker) are included for reference.}
    \label{fig:cubic_roots}
  \end{subfigure}\hfill
  \begin{subfigure}[b]{0.48\textwidth}
    \includegraphics[width=1\textwidth]{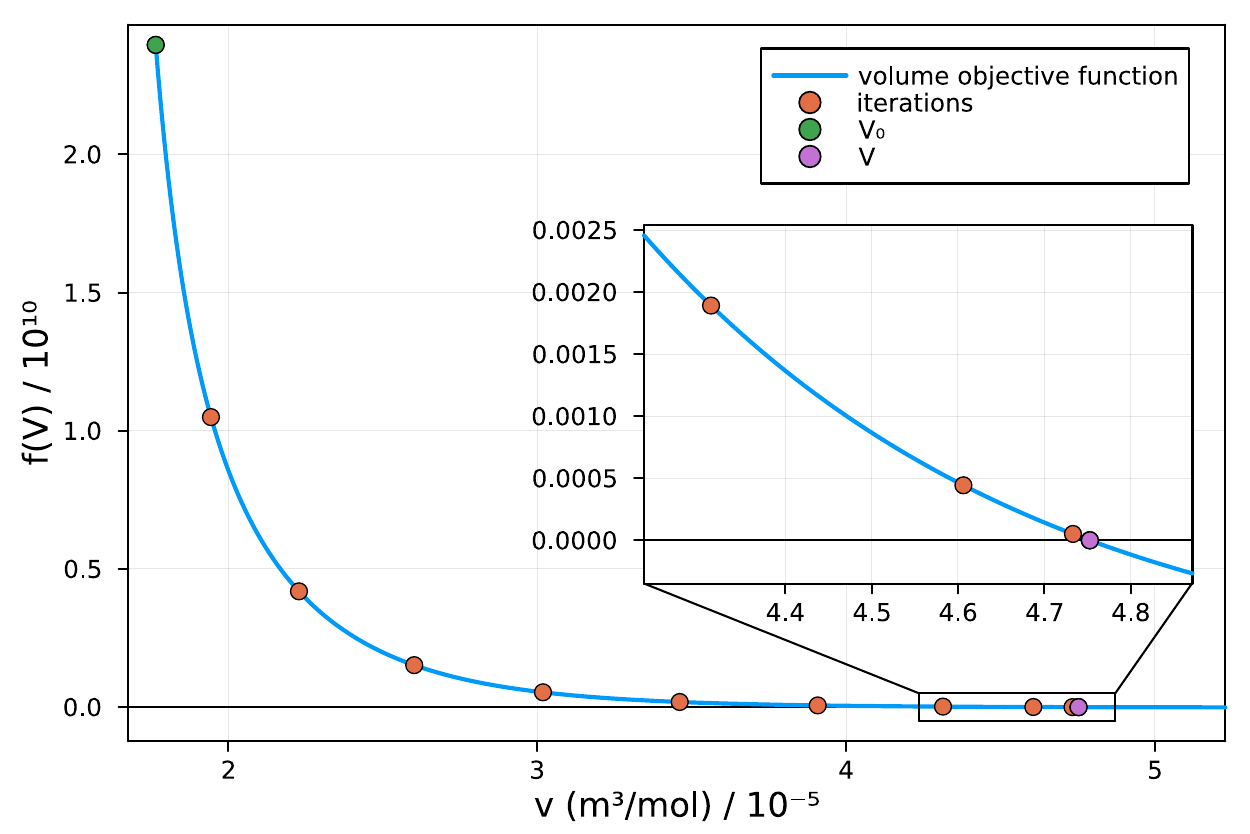}
    \caption{Convergence of the liquid volume using Newton's method from an initial guess (green) to the true solution (purple). The volume at each iteration is denoted by the orange markers. The continuous blue curve denotes objective function being solved.}
    \label{fig:convergence}
  \end{subfigure}
  \caption{Key illustrations used when discussing volume solvers within the Pluto notebooks.}
\end{figure*}

The simplest and perhaps most-useful method with which the students should be familiar with is that for obtaining the volume ($V$) of a system at a given temperature ($T_0$), pressure ($p_0$) and composition ($\mathbf{n}_0$):
\begin{equation}
    p(V,T_0,\mathbf{n}_0) = p_0\,.
    \label{eq:volume_solv}
\end{equation}
In the first instance, for simplicity of illustration, we focus on using the van der Waals equation of state, although we also highlight that any any cubic equation of state could be used.  The power of cubic equations of state becomes immediately apparent as equation \ref{eq:volume_solv} can be re-arranged to solve for the roots of a cubic polynomial:
\begin{equation}
    Z^3-\left(1+\frac{bp_0}{RT_0}\right)Z^2+\left(\frac{ap_0}{(RT_0)^2}\right)Z-\frac{abp_0^2}{(RT_0)^3}=0\,,
\end{equation}
where $Z=p_0V / (n_0RT_0)$ is the compressibility factor. The above equation can be solved analytically, obtaining all three roots. In the case where there is only one real root, no further considerations need to be made. However, if more than one real root is obtained, the most stable root is selected as that which corresponds to the lowest Gibbs free energy. The results from this procedure are demonstrated within a reactive $(p,V)$ diagram of carbon dioxide; students are invited to vary the temperature of the plotted isotherm (as shown in figure \ref{fig:cubic_roots}) and observe the resulting changes in the diagram. The metastable and unstable sections of the isotherm are included so as to illustrate the pressure construction (dashed black line) where the fluid will phase split.

We note that this problem only becomes more challenging when using equations of state with more-complex functional forms. We consider the SAFT equations of state, which must be solved numerically. However, instead of solving equation \ref{eq:volume_solv}, we remind students of the definition of the isothermal compressibility, $\beta_T$:
\begin{equation}
    \beta_T = -\frac{1}{V}\left(\frac{\partial V}{\partial p}\right)_T\,;
\end{equation}
if we assume that $\beta_T$ is constant, and integrate between states 1 and 2, we obtain:
\begin{equation}
    \ln{V_2} = \ln{V_1}+\beta_T(p_2-p_1)\,.
\end{equation}
If one were to set $p_2=p_0$, $\beta_T=\beta_T(V_1)$ and $p_1=p(V_1)$, then $\ln{V_2}$ would be our next-best guess for the volume, starting from some initial guess $V_1$, resulting in a recursive relationship:
\begin{equation}
    \ln{V_{i+1}}=\ln{V_i} + \beta_T(V_i)\cdot (p_0-p(V_i))\,.
\end{equation}
We point out that this iterative scheme is equivalent to solving equation \ref{eq:volume_solv} with respect to the logarithm of the volume. It is preferable to solve equation \ref{eq:volume_solv} with this approach as it offers greater numerical stability, particularly in the liquid phase. However, as with any iterative scheme, we need good initial guesses for each phase. For the vapour phase, the natural choice would simply be the volume of an ideal gas at the same conditions:
\begin{equation}
    V_0 = \frac{n_0RT_0}{p_0}\,.
\end{equation}
However, the situation for the liquid phase is more complicated. Thankfully, and the reason why we choose to use the SAFT equations as an example here, we can use physical intuition to obtain an initial guess for the liquid volume. We know that the minimum volume taken up by a species will be at least the summed total volume of the molecules. In the case of SAFT equations, the total volume taken up by one mole of a pure species is given by:
\begin{equation}
    V_\mathrm{min.} = \frac{\pi}{6}N_{\mathrm{A}}m\sigma^3\,,
\end{equation}
where $N_{\mathrm{A}}$ is Avogadro's number and the parameter $\sigma$ represents the diameter of each of the $m$ spherical segments comprising the model molecule. One can then set the initial guess of the volume phase to be some (low) multiple of this lower bound (we choose a multiple of 1.25). We do highlight that optimising this initial guess is of importance when accelerating the convergence of the algorithm (as demonstrated in figure \ref{fig:convergence}). Naturally, one must solve for both the vapour and liquid roots to then determine which of the two phases are stable.

We finish this exercise by pointing out that for more-complicated equations of state, such as the empirical equations of state, there are sometimes more than three roots at a given pressure and temperature. In such cases, it is important to ensure that all roots have been found before deciding which is most stable, as there is no guarantee that our initial guesses will lead to the two most-stable solutions.

\subsubsection{Saturation solvers for pure components}

One of the most-important classes of property for chemical engineers to obtain from equations of state is the saturation properties of pure components. Solving for the saturation conditions at a given temperature is more complicated than the volume solvers mentioned previously as it involves solving for the volume ($v^\mathrm{liq.}$ and $v^\mathrm{vap.}$) of each phase using the conditions for phase equilibrium between two phases:
\begin{align}
    p^\mathrm{liq.}-p^\mathrm{vap.} &= 0\,,\nonumber\\
    \mu^\mathrm{liq.}-\mu^\mathrm{vap.} &= 0\,,\label{eq:phaseq}
\end{align}
where the superscripts liq. and vap. denote properties relating to the liquid and vapour phases, respectively. Inherently, this problem amounts to simply using the Newton--Raphson method to converge the volumes, allowing us to illustrate important concepts when solving such problems. For example, we re-iterate the importance of initial guesses, using the approaches we developed in the volume solver, or using empirical correlations. In addition, we also highlight the value of re-scaling the variables such that the numerical solver is more stable (including solving for the logarithm of the volumes rather than the absolute values, and normalising equation \ref{eq:phaseq} by the characteristic length scales of pressure and energy).

Students will now be able to trace the saturation curve. However, we point out that, for some equations of state, such as SAFT-type equations, the end-point of the saturation curve, i.e., the critical point, isn't known \textit{a priori}; this must be solved for numerically. We remind students of the definition for the critical point of a pure substance:
\begin{align}
\left(\frac{\partial p(v,T)}{\partial v}\right)_{T} &= 0\,,\nonumber\\
\left(\frac{\partial^2 p(v,T)}{\partial v^2}\right)_{T} &= 0\,.
\end{align}
Students can re-use the tricks we highlighted in relation to the saturation solver to solve the above system of equations. As a result, the student can now trace the entire vapour--liquid envelope.

\subsubsection{Chemical Stability Analysis}

\begin{figure*}[!ht]
\centering
\begin{subfigure}[b]{0.48\textwidth}
    \includegraphics[width=1\textwidth]{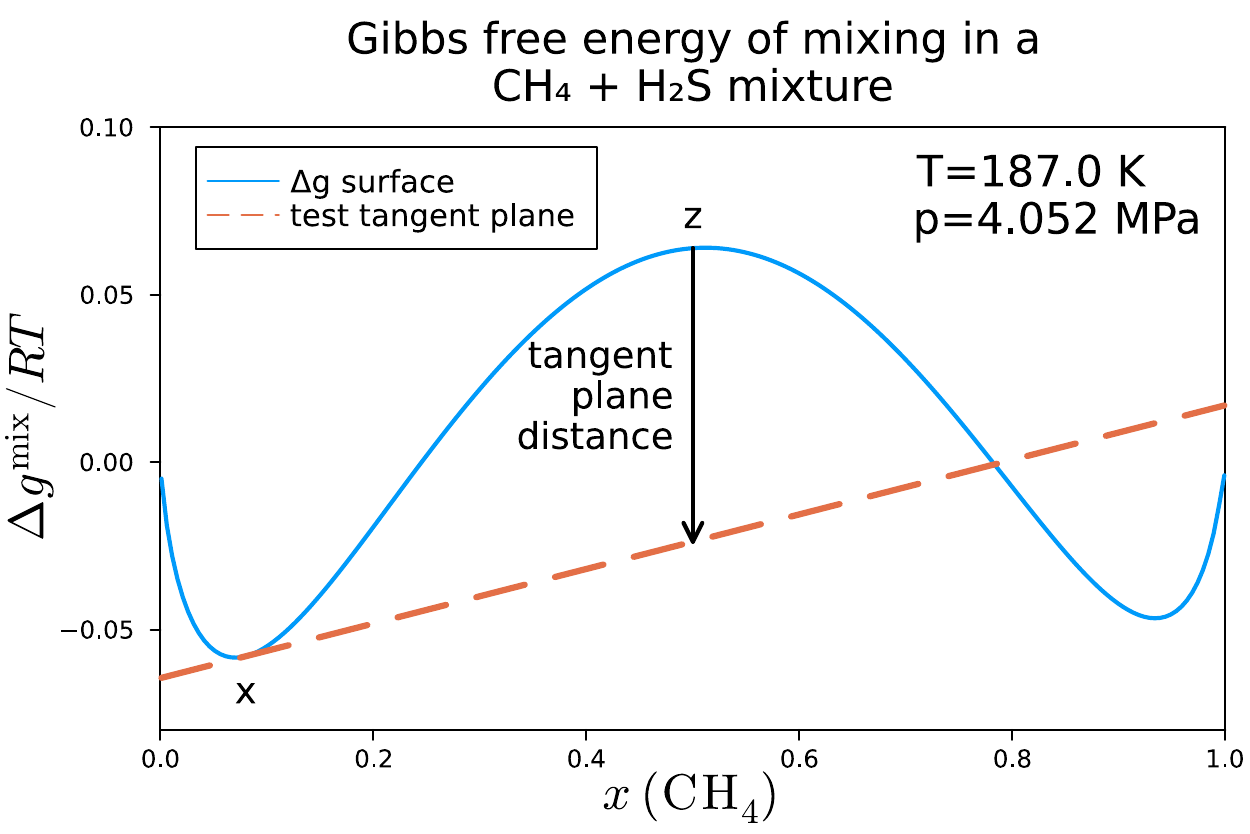}
    \caption{Definition of the Gibbs tangent plane distance (black line). The continuous blue curve denotes the Gibbs free energy of mixing and the dashed orange line denotes the test tangent plane.}
    \label{fig:tpd}
  \end{subfigure}\hfill
  \begin{subfigure}[b]{0.48\textwidth}
    \includegraphics[width=1\textwidth]{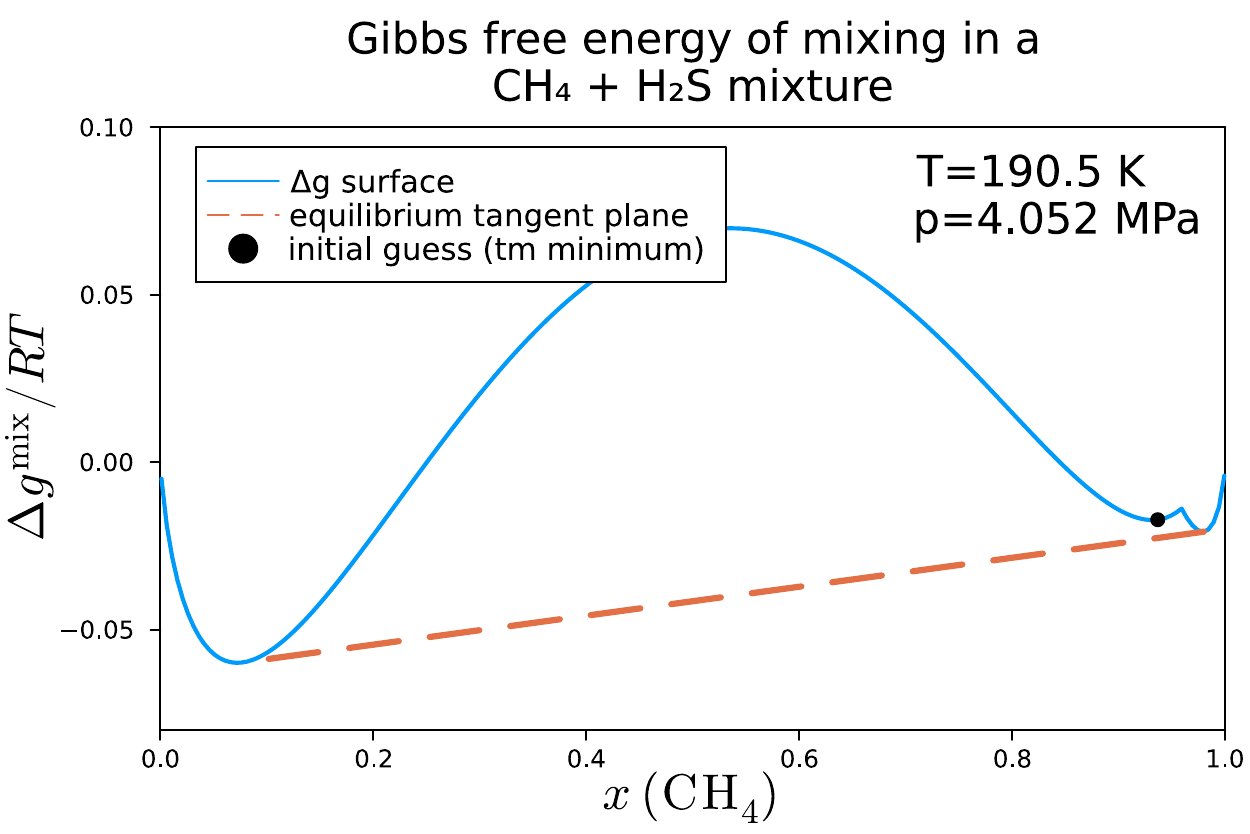}
    \caption{Example conditions where chemical-stability analysis may fail. Local optimisers would not recognise that initial conditions denoted by the black dot are unstable. The dashed orange curve here denotes the co-existence tangent plane.}
    \label{fig:vlle}
  \end{subfigure}
  \caption{Key illustrations used when discussing chemical stability analysis within the Pluto notebooks.}
  \label{fig:pure_phase_diagrams}
\end{figure*}

While the previous exercises pertained to single-component systems, for most applications, chemical engineers will be dealing with mixture systems. The phase space of mixtures becomes more complex as we introduce wide regions where the system may exist in two or more phases. As such, if we wish to study the properties of a system at given set of conditions ($p_0,T_0,\mathbf{z}_0$), we must first answer the question: ``Does a phase split occur?". This problem is non-trivial and involves carrying out chemical-stability analysis. The standard approach was developed by Michelsen\cite{michelsen_isothermal_1982} and involves minimising what is referred to as the Gibbs tangent plane distance ($TPD$):
\begin{equation}
    TPD(\mathbf{x}) = \sum_i x_i(\mu_i(\mathbf{x})-\mu_i(\mathbf{z}_0))\,,
\end{equation}
where $x_i$ is the mole fraction of species $i$ in a candidate phase. The function is demonstrated visually in figure \ref{fig:tpd}. If $TPD<0$, then we know that a phase split will occur. 

Once again, we take the students through the process of re-defining the problem in a way that is more numerically stable. We also take the opportunity to illustrate that, unless a global optimiser is used, there is a chance that the chemical-stability analysis might fail. As shown in figure \ref{fig:vlle}, if a local optimiser is used, conditions near the black marker would not be seen as unstable as, despite being locally stable, there exists a tangent plane with a lower Gibbs free energy (denoted by the dashed line in the figure).

\subsubsection{Isothermal Flash Problem}

Having now established when a phase split will occur, the next step is to determine what the various phases will be. Balancing both simplicity and rigour, we choose to focus solely on the isothermal flash problem, the backbone of multiple vital processes in chemical engineering. 

To build up the necessary tools to perform the flash calculations, we derive the relationship between $K$-factors and fugacity coefficients from the definition for chemical equilibrium. From here, we impose the mass-balance equation. As a result, we obtain the venerable Rachford--Rice\cite{rachford_procedure_1952} equation:
\begin{equation}
    f(\beta) = \sum_i \frac{(K_i-1)z_i}{1+\beta(K_i-1)}=0\,,
    \label{eq:rachford_rice}
\end{equation}
where $\beta$ is the phase fraction and $K_i$ is the $K$-factor for species $i$. Solving equation \ref{eq:rachford_rice} is at the heart of almost all flash algorithms. As such, we task the students with solving the Rachford--Rice equation. In introducing the problem, we highlight one of the benefits of this equation: it is monotonically decreasing. The implementation of the solver is left to the students to fill-in, within the sample code. The exercise is presented the same way as the previous exercise shown in figure \ref{fig:cubic_example}.

Students will thus be able to obtain the phase fraction, for a given set of $K$-factors, representing the first iteration of the flash algorithm. For most real systems, $K$-factors depend on the composition in each phase. Accordingly, once we've solved for the phase fraction, we must update the $K$-factors and verify if the composition of each phase has changed significantly. If so, we must solve the Rachford--Rice equation again with the updated $K$-factors, leading to an iterative scheme. 

\begin{figure*}[!ht]
\centering
\begin{subfigure}[b]{0.48\textwidth}
    \includegraphics[width=1\textwidth]{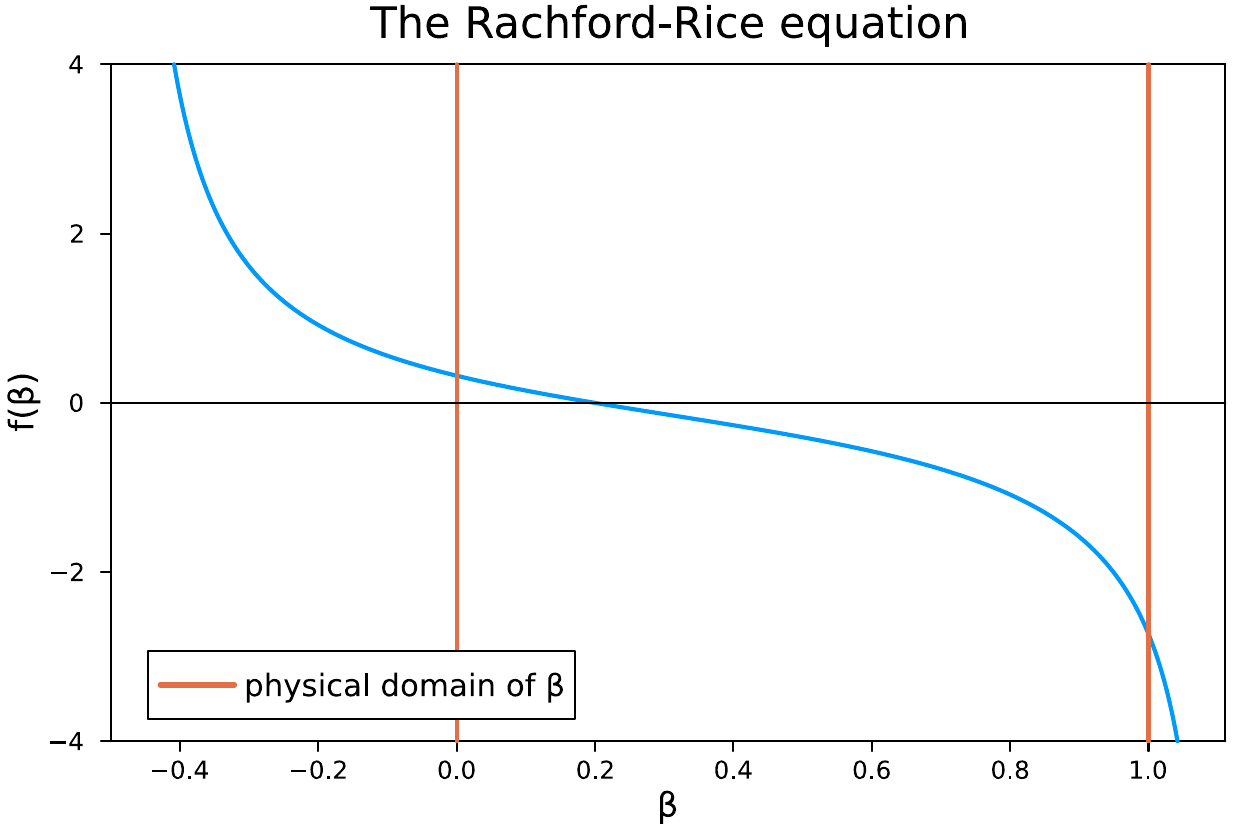}
    \caption{Rachford--Rice equation as a function of phase fraction.}
    \label{fig:RR}
  \end{subfigure}\hfill
  \begin{subfigure}[b]{0.48\textwidth}
    \includegraphics[width=1\textwidth]{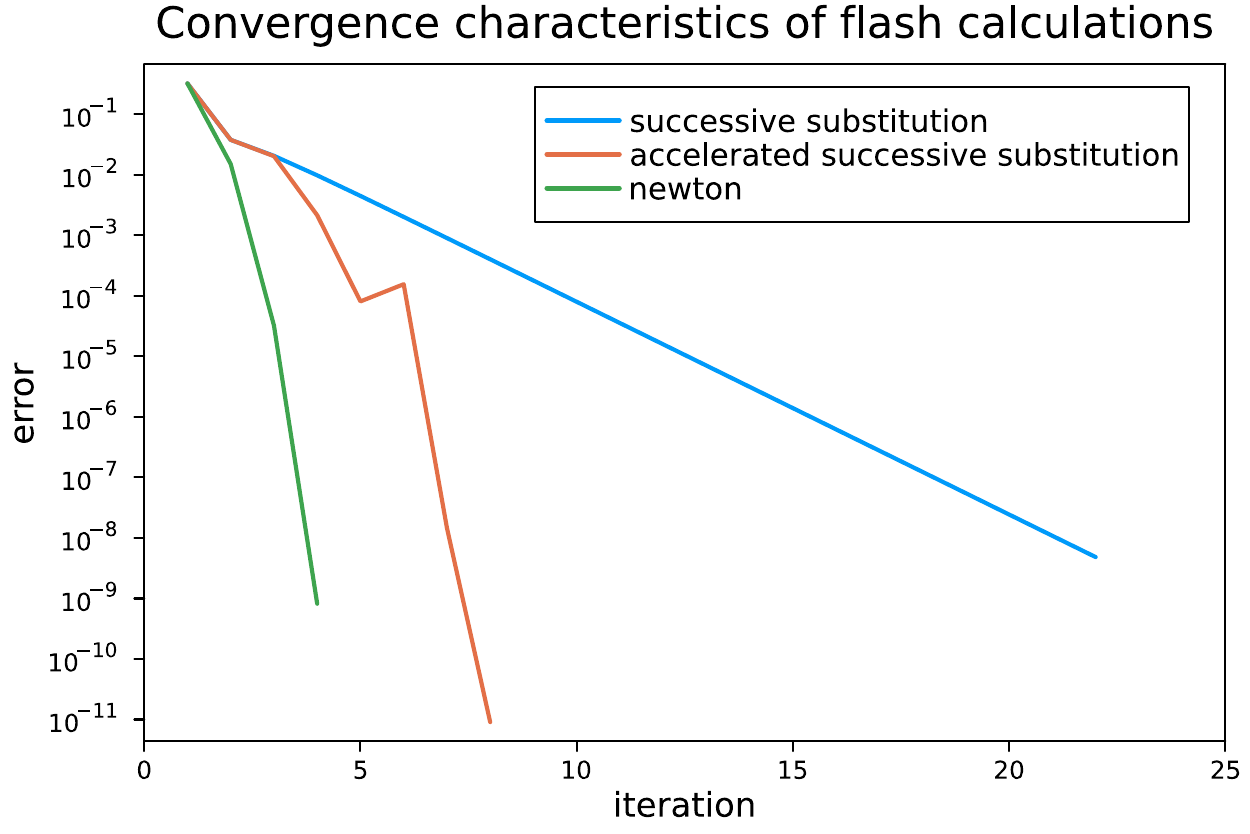}
    \caption{Convergence of flash algorithm using different iterative methods.}
    \label{fig:flash_convergence}
  \end{subfigure}
  
  \caption{Key illustrations used when discussing chemical stability analysis within the Pluto notebooks.}
  \label{fig:flash}
\end{figure*}

While the implementation we present uses a simple successive-substitution iterative scheme, we highlight that more-advanced flash algorithms, such as the Michelsen flash algorithm\cite{michelsen_isothermal_1982}, use methods such as Newton--Raphson or Anderson Acceleration to improve the convergence (shown visually in figure \ref{fig:flash_convergence}). We highlight that, while these algorithms may require fewer iterations to convergence, it is important to consider the computational cost of each iteration as a Newton--Raphson step can be significantly more costly than an Anderson Acceleration step.

By this point, students will have a strong foundation to flash algorithms, allowing them to comfortably explore the literature on this topic at their own leisure. We provide multiple references on more-advanced flash algorithms for them to explore independently. 

\section{Course pilot: results and feedback}
\label{sect:results}

A remote pilot run of the course took place in the summer of 2022 with 16 participants. As the course is intended for students who have already had at least one introductory thermodynamics class, both undergraduate (beyond second year) and graduate students were invited to participate. A pre-course and a post-course survey were delivered. Both surveys were vetted and approved by the Ethics Research Committee at Imperial College London and were anonymised with the assistance of the \textit{StudentShapers} programme\cite{studentshapers_nodate}. Some questions presented in the second survey were repeats of questions presented in the first, so as to evaluate the change in student self-efficacy. The pre-course survey contained questions designed to determine students' general coding proficiency prior to taking the course. The post-course survey contained more questions related to the students' experience and thoughts on the course. Open-ended questions were also added to the post-course survey in order to allow students to fully express any feedback they might have.

\begin{figure}
    \centering
    \includegraphics[width=0.5\textwidth]{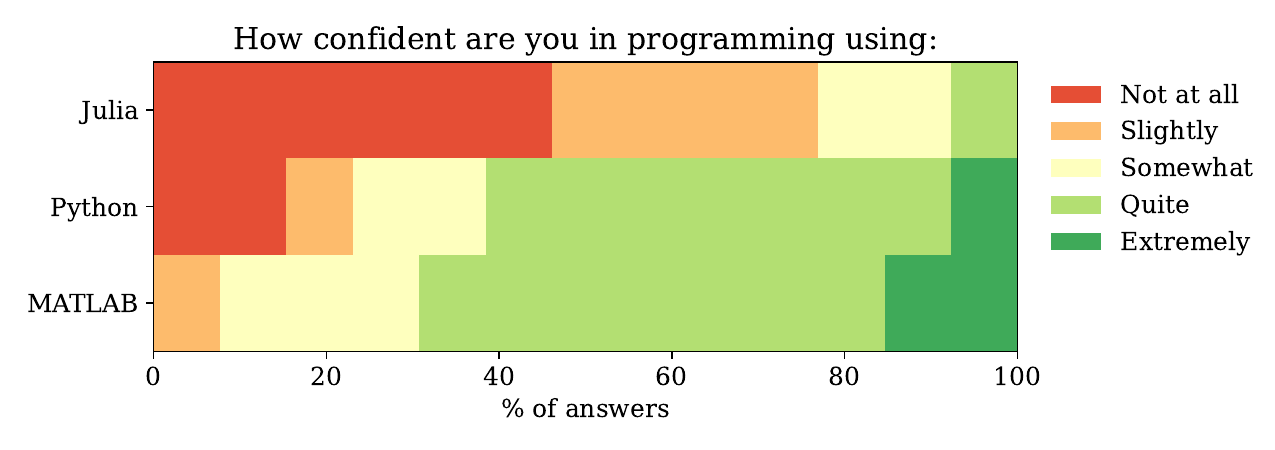}
    \caption{Summary of student response when asked to qualify their confidence in programming using different languages.}
    \label{fig:coding_lang}
\end{figure}

\begin{figure}
    \centering
    \includegraphics[width=0.5\textwidth]{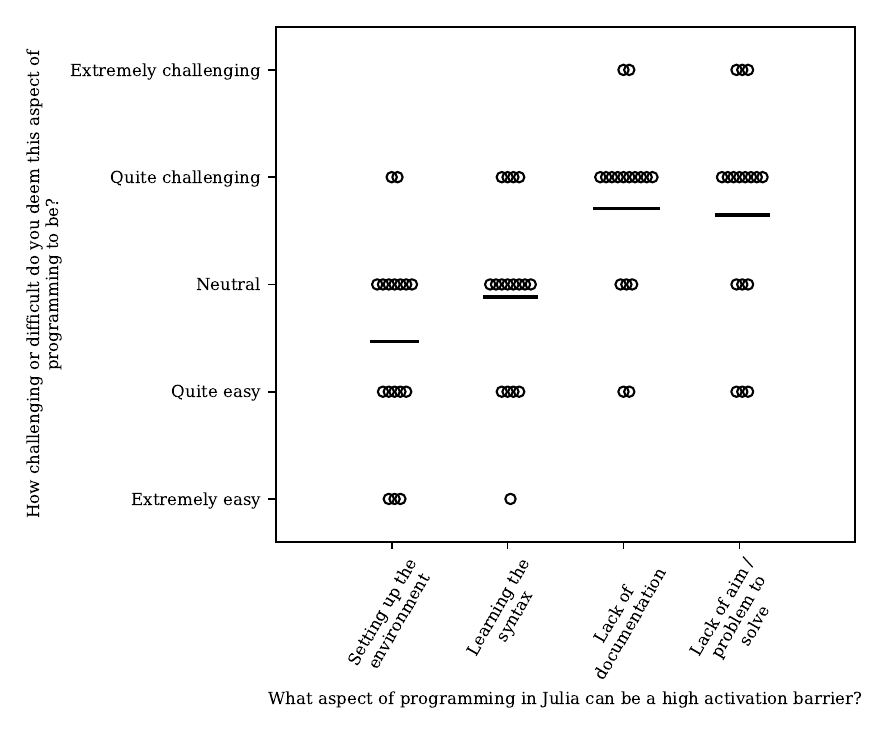}
    \caption{Student rating to aspects of programming in the Julia language which present an activation barrier. Symbols represent individual response and solid lines represent the mean response.}
    \label{fig:activation_barrier}
\end{figure}

\subsection{Introduction and code familiarisation}

One of the main factors motivating the decision to use the Julia language and Pluto notebooks for this course is to reduce the friction students encounter when coding and/or using new software, which had been an issue in prior work \cite{inguva_introducing_2021}. When asked at the start of the course how confident students felt in programming with the MATLAB, Python, and Julia languages, the majority were either not at all or only slightly confident with using Julia, as shown in figure \ref{fig:coding_lang}. In comparison, students were mostly quite or extremely confident in using MATLAB and/or Python. The relative lack of confidence in using Julia is not surprising considering that it is a newer language and many students would not have encountered its use. In anticipation of students' lack of familiarity, a detailed set of instructions for installing and using the code environment was prepared and made available on the Github repository. We also offered a one-hour diagnostic and troubleshooting session prior to the course -- which, surprisingly, no student attended. In contrast to our previous experience with Python, students mostly did not find either setting up the environment or learning the syntax of Julia to be a significant activation barrier (see figure \ref{fig:coding_lang}). The latter may not be surprising as the syntax used in MATLAB (the language which most students were familiar) is quite similar to that used in Julia. The former speaks to the simplicity with which the Julia package manager, and Pluto, abstract away the complexities typically experienced when setting up a coding environment, demonstrating the benefits of using Julia in an educational setting. One student commented on the similarity between MATLAB and Julia, as well as their experience with the Pluto notebooks:

\textit{``I enjoyed the interactive notebooks which allowed for a better learning experience. Reminded me of MATLAB on-ramp course, which was a good introduction/helper course.''}

The lack of documentation, which was not a significant a issue with Python, has become more of a barrier. This is primarily due to how comparatively young Julia is as a language. As a result, packages in Julia often have comparatively less comprehensive documentation and support than their Python/MATLAB equivalents. Another common source of help for coders are online forums like StackOverflow \cite{bhasin_student_2021,dondio_is_2019} and blog posts are also similarly less developed for Julia. Nonetheless, we believe that this issue will improve over time considering the rapid development and adoption of Julia by many groups such as educators and researchers. Also highlighted in figure \ref{fig:activation_barrier} is another limitation of introducing students to coding, which has persisted from our prior work, is the lack of aim / problems to solve. This is something we hope to have resolved with this and future courses we develop. 

\subsection{Content and structure}

\begin{figure}
    \centering
    \includegraphics[width=0.5\textwidth]{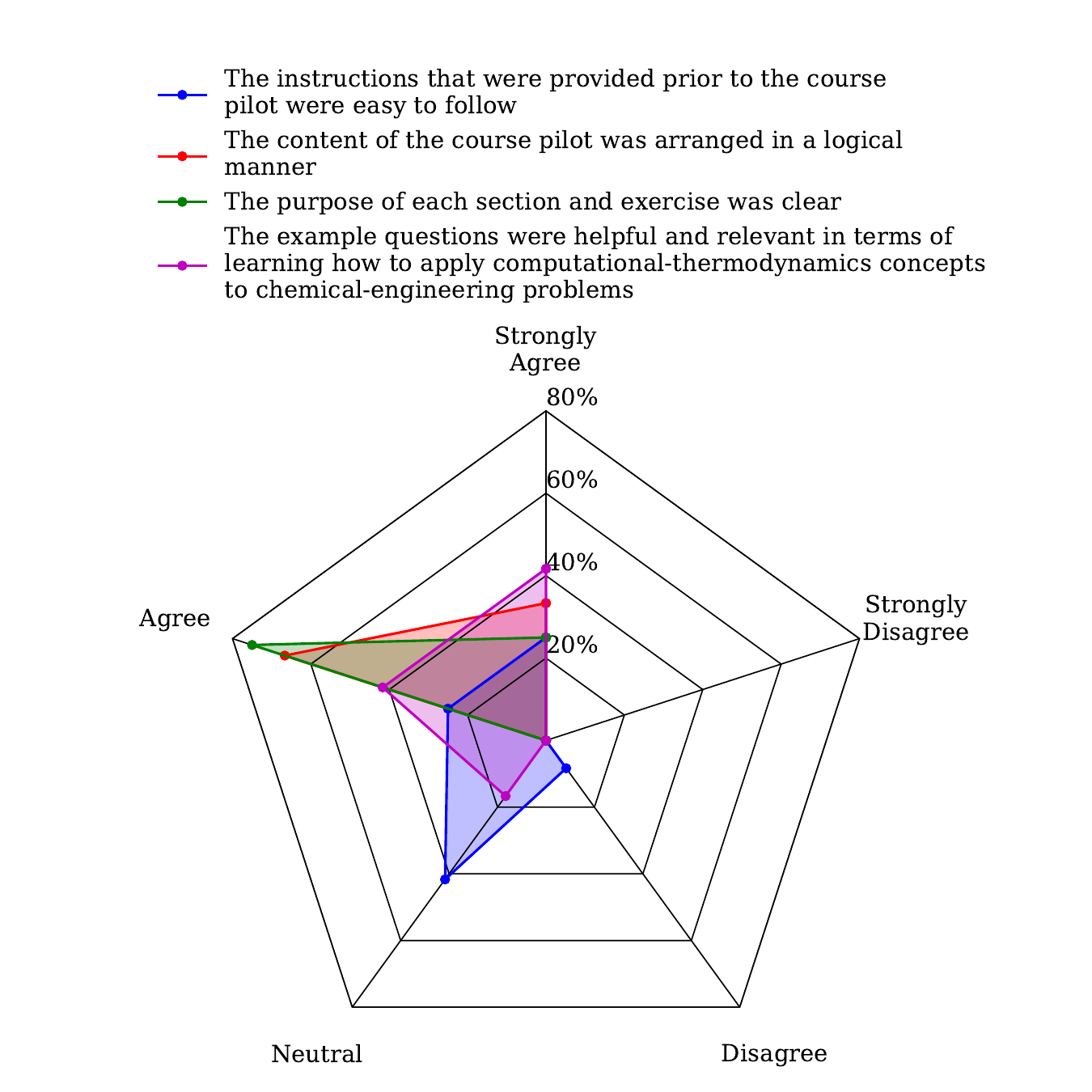}
    \caption{Student responses to various aspects of the course delivery.}
    \label{fig:course_delivery}
\end{figure}

\begin{figure*}[!ht]
\centering
\begin{subfigure}[b]{0.49\textwidth}
    \includegraphics[width=1.\textwidth]{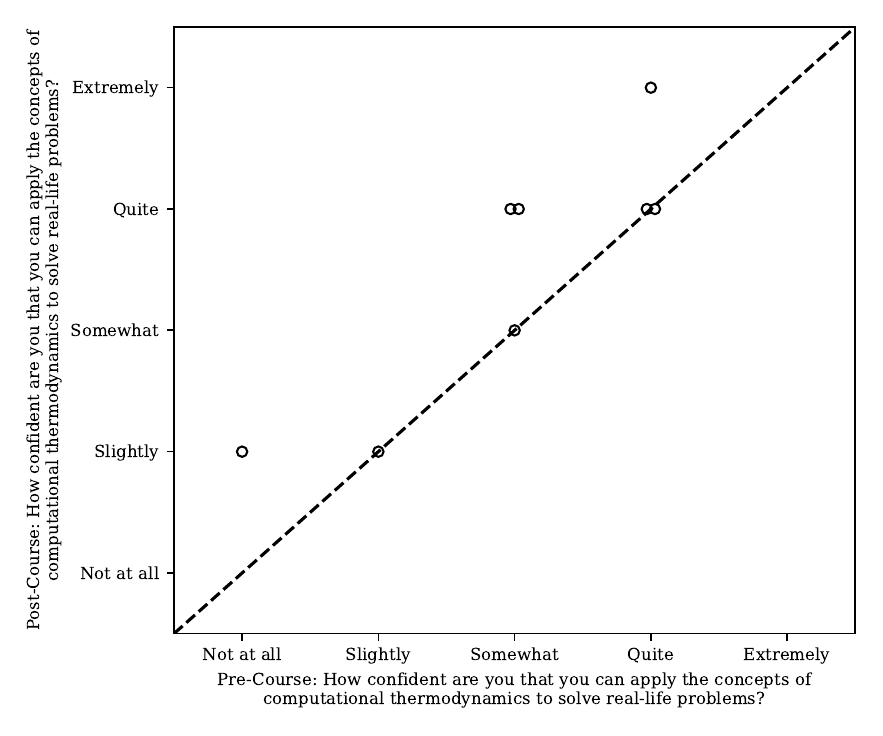}
    \caption{Student responses to their self-perceived confidence to apply concepts of computational thermodynamics in real life.}
    \label{fig:application}
  \end{subfigure}\hfill
  \begin{subfigure}[b]{0.49\textwidth}
    \includegraphics[width=1.\textwidth]{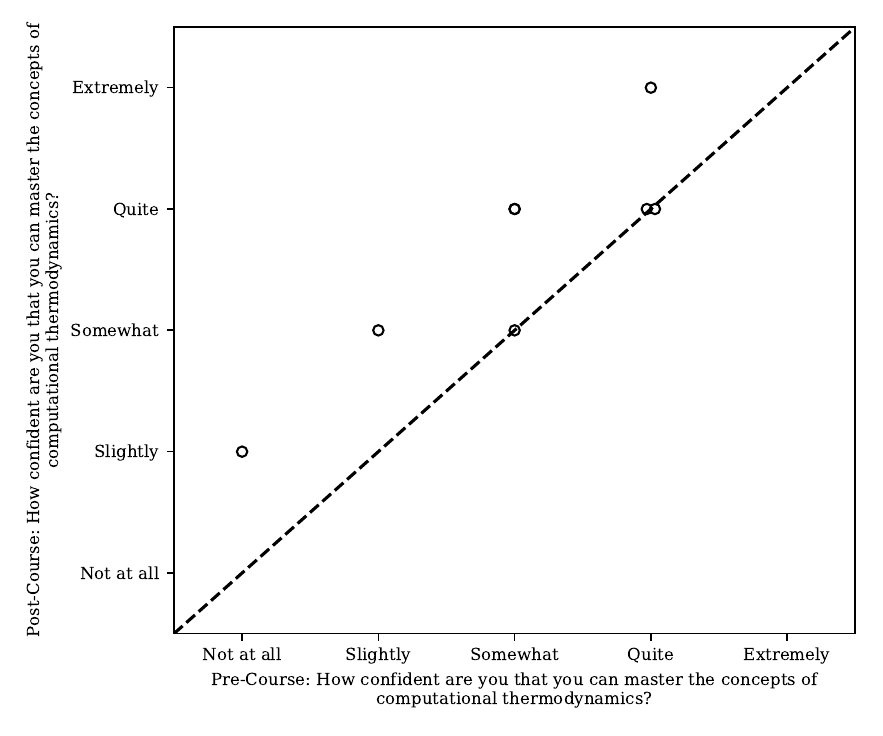}
    \caption{Student response to their self-perceived confidence to master concepts of computational thermodynamics.}
    \label{fig:mastering}
  \end{subfigure}
  
  \caption{`Before-and-after" student responses related to their self-perceived confidence in aspects of computational thermodynamics before and after the course.}
  \label{fig:improve}
\end{figure*}

Overall, the feedback for the course content and structure was quite positive. This is summarised in figure \ref{fig:course_delivery}. As we can see, the main area of improvement is the instructions provided prior to the course, which came in the form of an e-mail providing information and links to the online repository where students could obtain more details on setting-up the coding environment. Had the course been held in-person, perhaps it would have been more instructive to formally set-up an in-person session for this. However, we were restricted by the remote delivery of the course. Nevertheless, as demonstrated in the previous section, this did not seem to be a significant issue as most students successfully set-up the coding environment and became familiarised with the syntax quickly, despite over half ($N=8$) not using code frequently.

As evident from figure \ref{fig:course_delivery}, the arrangement of content and purpose of each section was very well received, highlighting the value of using progression as a guiding principle in the course development\cite{xie_accelerating_2020}. We are particularly happy that most students agreed that the examples given were relevant to chemical-engineering problems. To this end, one of the more-emblematic responses we received from students after being asked which part of the course they enjoyed most:

\textit{``The history of EoS as it really makes you understand how far we have come.''}

This was indeed one of the main motivations for the development of this course: bringing students up-to-speed with the modern class of equations of state available today\cite{de_hemptinne_view_2022}. This is made even more apparent in figure \ref{fig:improve} where it is illustrated that, when students were asked whether or not the course improved their confidence with regards to mastering and applying the concepts of computational thermodynamics, there was a near-total increase. Based on one student's feedback:

\textit{``...the code template made it easy to actually solve the exercises despite lacking the necessary knowledge to set up the problem in code.'',}

It appears that the integration of code templates and explanatory text was effective in helping students develop a good understanding of computational thermodynamics, despite not necessarily having a strong computational background. Interestingly, one of the pieces of feedback we received from students was that we \textit{``Could have more activity sessions..."}. The number of activity sessions (reactive exercises) was restricted to just two since we intended to develop an accelerated introduction to computational thermodynamics.
% we believed that having more exercises would increase the total length of the course. 
However, to compensate for this, we embedded many examples within the Pluto notebooks, so that students might have the opportunity to examine the code after the course, by which time they would have a stronger computational-thermodynamics background. Nevertheless, clearly the students felt that, when interacting with code, exercises were more effective as they felt more comfortable when under the supervision of the teaching staff. 

Another improvement that was suggested related to the pace of the course. With three, three-hour long lectures (including two ten-minute breaks), one per day, it was apparent each day that by the final hour, students had a harder time concentrating. As such, in future implementations of the course, it would be advised to extend the duration to five or six sessions in order to spread out the material and give more opportunities for questions and interaction with the Pluto notebooks.

\section{Conclusions}
\label{sect:conclusion}

We have developed and rolled-out an introductory course on computational thermodynamics for both undergraduate and graduate students using the open-source thermo-fluids package, Clapeyron.jl, and the reactive notebooks provided by Pluto.jl. Through the course, students have been able to develop a sufficiently high level of competency in the use of a state-of-the-art thermodynamics software package and a familiarity with concepts and methods in computational thermodynamics. This result is demonstrated by the overall increase in students' self-efficacy and confidence. The course thus helps position students to become informed users and also obtain a sufficient foundation to engage with the thermodynamics literature. Two key components enable such rapid and effective learning: first, the use of progression as a guiding philosophy in course design where content and exercises are structured such that content and exercises progressively reach the required level of complexity and sophistication; second, the use of the Julia language and reactive notebooks. 

In a previous course developed by the authors \cite{inguva_introducing_2021}, we used template Python scripts for code implementation. While the use of the Python language presented no issues due to its familiar syntax, setting up the computational environment was a ``pain point" for students and the use of static scripts limited the opportunities for student engagement during the course. In this course, the use of the Julia language and Pluto notebooks successfully addressed these two issues while also not presenting any significant challenges from the perspective of introducing a new coding language. Setup was much more straightforward, even with none of the participants attending the pre-course trouble-shooting session.  The use of reactive notebooks also enabled integration of course content with code in an environment that facilitates student engagement and exploration. 

This course represents an important first step in addressing a pertinent gap in advanced thermodynamics instruction. Providing students with the opportunity to understand how thermodynamic theory is employed in modern applications is important for their own practice and also to accelerate the promulgation of developments in thermodynamics to the rest of the scientific and engineering community. The modular nature of the course content enables self-directed learners and other educators to pick and choose components that are relevant to them and also to easily develop new material for their own use/courses. However, due to the comparatively small number of students who participated in the pilot study, further evaluation is important for evaluating whether the proposed course structure and delivery format will be effective for a larger cohort or if a similar quality of learning experience can be achieved when incorporating parts of the course into a larger thermodynamics module.

\section*{Availability of Code}
All the course notes and code can be found at the following repository: \url{https://github.com/ClapeyronThermo/introduction-to-computational-thermodynamics}

\section*{Acknowledgments}
The authors would like to thank the students who participated in this course and provided their valuable feedback. This project was supported by funding from StudentShapers (Imperial College London) to enable partnership with students.

\clearpage
\printbibliography

\end{document}